\documentclass[aps,prd,superscriptaddress,showpacs,preprint]{revtex4}
\usepackage{graphicx, bm}
\begin{document}
\draft
\title{Bounds on the anomalous $HZ\gamma$ vertex arising from the\\
       process $e^+e^-\to \tau^+ \tau^- \gamma$}

\author{A. Guti\'errez-Rodr\'{\i}guez}
\affiliation{\small Facultad de F\'{\i}sica, Universidad Aut\'onoma de Zacatecas\\
         Apartado Postal C-580, 98060 Zacatecas, M\'exico.\\}

\author{J. Monta\~no}
\affiliation{\small Departamento de F\'{\i}sica, CINVESTAV.\\
             Apartado Postal 14-740, 07000, M\'exico D.F., M\'exico.}

\author{M. A. P\'erez}
\affiliation{\small Departamento de F\'{\i}sica, CINVESTAV.\\
             Apartado Postal 14-740, 07000, M\'exico D.F., M\'exico.}

\date{\today}

\begin{abstract}

We obtain limits on the anomalous coupling $HZ\gamma$ through data
published by the L3 Collaboration on the process $e^+e^-\to
\tau^+\tau^- \gamma$. Our analysis leads to bounds on this
coupling of order $10^{-2}$, for an intermediate mass Higgs boson
$115 < M_H < 145$ $GeV$, two orders of magnitude above the
Standard Model prediction.
\end{abstract}

\pacs{14.60.Fg, 12.60.-i
\vspace*{2cm}\noindent  E-mail: $^{1}$alexgu@fisica.uaz.edu.mx,
$^{2}$mperez@fis.cinvestav.mx}

\vspace{5mm}

\maketitle


\section{Introduction}

In the Standard Model (SM) of electroweak interactions there are
no couplings at the tree level among three neutral bosons
involving photons, such as $HZ\gamma$. These couplings only appear
at the one-loop level through fermion and charged vector bosons
\cite{Ellis,Gunion,Cotti}. In the SM it is dominated by $W$ gauge
boson and top quark loops and the branching ratio for the decay
mode $H \to Z\gamma$ reaches its maximum value of order $10^{-3}$
for an intermediate-mass Higgs boson ($115 < M_H < 140$ $GeV$)
\cite{Cotti}. The study of this vertex has attracted much
attention because its strength can be sensitive to scales beyond
the SM. The interest in this type of couplings thus lies in the
additional contributions that may appear in extensions of the SM.
This is the case, for example, for contributions coming from new
charged scalar and vector bosons in Left-Right (L-R) symmetric
gauge models \cite{Martinez}, or Two Higgs Doublet Models (THDM)
\cite{Diaz-Cruz,Bates}, as well as charginos and neutralinos in
the Minimal Supersymmetric Standard Model (MSSM)
\cite{Diaz-Cruz,Bates}. The SM and L-R symmetric models predict an
anomalous $HZ\gamma$ vertex of order $10^{-4}$
\cite{Ellis,Gunion,Cotti}, the MSSM may induce a suppression
effect \cite{Diaz-Cruz,Bates} but an effective Lagrangian approach
leaves room for an enhancement effect \cite{Diaz-Cruz,Hernandez}.
It has been found also that the QCD corrections in the SM are well
under control \cite{Spira}. A measurement of this vertex thus may
be used to distinguish among theories beyond the SM.

The sensitivity to the $HZ\gamma$ vertex has been studied in
processes like $e^-\gamma \to e^-H$ and $e^+e^- \to H\gamma$
\cite{Cotti,Gounaris,Hagiwara}, rare $Z$ and $H$ decays
\cite{Li,Perez,Hankele}, $pp$ collisions via the basic interaction
$qq \to qqH$ \cite{Hankele} and the annihilation process $e^+e^-
\to HZ$ \cite{Hagiwara,Rindani,Killian}. It has been found that
the latter reaction with polarized beams may lead to the best
sensitivity to the $HZ\gamma$ vertex \cite{Rindani} while an
anomalous $HZ\gamma$ coupling may enhance partial Higgs decays
widths by several orders of magnitude that would lead to
measurable effects in Higgs signals at the LHC \cite{Hankele}.

The general aim of the present paper is to obtain limits on the
$HZ\gamma$ vertex coming from the LEP-I data on the reaction
$e^+e^- \to \tau^+\tau^-\gamma$ \cite{L3,OPAL}. We will find
limits of order $10^{-2}$, which are better by an order of
magnitude than the bounds obtained from the known limits on the
partial decay widths of the $Z$ boson \cite{Hankele}, but still
two orders of magnitude above the SM prediction
\cite{Cotti,Diaz-Cruz}. The L3 collaboration has obtained also
limits on the $HZ\gamma$ vertex using now LEP-II data for events
with photons and a $Z$ vector boson in the final state
\cite{Achard}. In this case they have used an analysis that
involves the Higgs boson decay modes $H \to \gamma\gamma$,
$Z\gamma$. We have found that our analysis with a tau-lepton pair
in the final state induces more stringent limits on the $HZ\gamma$
vertex.

In Fig. 1, we show the Feynman diagrams which give rise to the
process $e^{+}e^{-}\rightarrow \tau^+ \tau^- \gamma$ in the SM at
tree level and with the anomalous $HZ\gamma$ vertex when the $Z$
vector boson is produced on mass-shell. We do not include the
contribution coming from initial photon bremsstrahlung because the
L3 data considered the appropriate energy cuts to reduce this
contribution.

The paper is organized as follows. In Section II we present the
calculation of the respective cross section and in Section III we
presented our results and conclusions.

\section{CROSS SECTIONS}

The anomalous $V^\mu_1(p_1)-V^\nu_2(p_2)-H(p_H)$ vertex function
is given by \cite{Hernandez,Hagiwara}

\begin{equation}
\Gamma^{HV_1V_2}_{\mu\nu} (p_H, p_1, p_2)= g_ZM^2_Z \biggl[
h^{V_1V_2}_1g_{\mu\nu} +
\frac{h^{V_1V_2}_2}{M^2_Z}p_{2\mu}p_{1\nu}\biggr],
\end{equation}

\noindent where $M_Z$ is the $Z$ boson mass,
$g_Z=e/\sin\theta_W\cos\theta_W$ and $V_1, V_2$ can be $(V_1V_2)=
(ZZ), (Z\gamma), (\gamma Z), (\gamma \gamma), (W^+W^-)$ or
$(W^+W^-)$. The coefficients $h^{V_1V_2}_i$ are

\begin{eqnarray}
h^{Z\gamma}_{1}(p_{1},p_{2}) & = &
\frac{p^{2}_{1}+p^{2}_{2}-m^{2}_{H}}{m^{2}_{Z}}c_{2 Z \gamma}
-\frac{p^{2}_{1}-p^{2}_{2}-m^{2}_{H}}{m^{2}_{Z}}c_{3 Z \gamma},\\
h^{Z\gamma}_{2}(p_{1},p_{2}) & = & 2(c_{2 Z \gamma}-c_{3 Z
\gamma}),\end{eqnarray}

\noindent for the $HZ\gamma$ couplings. The coefficients $c_{2 Z
\gamma}$ and $c_{3 Z \gamma}$ are given explicitly in Eqs. (12d)
and (12h) of Ref. \cite{Hagiwara}.

In the present study we have considered only CP-conserving
$HZ\gamma$ couplings but our results can be applied also for the
CP-violating coupling. Since the L3 Collaboration \cite{Achard}
has used the $H \to Z\gamma$ decay rate in order to get its limits
on the $HZ\gamma$ coupling, we present below the expression for
its decay width

\begin{equation}
\Gamma(H\to Z\gamma)=
\frac{g_Z^2}{8\pi}\frac{m_Z^2\Bigl(m_H^2-m_Z^2\Bigr)}{m_H^3}\Bigl(h_1^{Z\gamma}\Bigr)^2,
\end{equation}

\noindent where we have used the fact that for on-shell particles,
only one of the form factors given in Eq. (1) contribute to the
decay width \cite{Diaz-Cruz}.

The expression for the respective cross section, that includes the
SM and the $HZ\gamma$ vertex contributions shown in Fig. 1, is
given by

\begin{eqnarray}
\sigma(e^+e^-\to \tau\bar\tau\gamma)&=& \int
\frac{\alpha^3}{96}\biggl[3m^2_\tau C_1(x_W) \Bigl[F_1(s,
E_\gamma,
\cos\theta_\gamma) (h^{Z\gamma}_{1})^2+ F_2(s, E_\gamma, \cos\theta_\gamma)(h^{Z\gamma}_{2})^2 \Bigr]\nonumber\\
&+& m^2_\tau C_2(x_W)\Bigl[F_3(s, E_\gamma, \cos\theta_\gamma)
h^{Z\gamma}_{1}+F_4(s, E_\gamma, \cos\theta_\gamma)
h^{Z\gamma}_{2}\Bigr]\\
&+& C_3(x_W)F_5(s, E_\gamma, \cos\theta_\gamma)\biggr]E_\gamma
dE_\gamma d\cos\theta_\gamma,\nonumber
\end{eqnarray}

\noindent where $\alpha=e^2/4\pi$ is the electromagnetic coupling,
$E_\gamma$ and $\cos\theta_\gamma$ are the energy and scattering
angle of the photon and the $C_{1, 2, 3}$ coefficients label the
contributions arising from the $HZ\gamma$, interference, and SM
amplitudes, respectively. The kinematics is contained in the
functions

\begin{eqnarray}
F_1(s, E_\gamma, \cos\theta_\gamma)&\equiv&
\frac{\Bigl(\frac{1}{2}s -
\sqrt{s}E_\gamma - 2m^2_\tau\Bigr)}{[(s-M^2_Z)^2+M^2_Z\Gamma^2_Z](s+2\sqrt{s}E_\gamma-M_H^2)^2},\nonumber\\
F_2(s, E_\gamma, \cos\theta_\gamma)&\equiv&
\frac{\Bigl(\frac{1}{6}E^2_\gamma
-\frac{1}{3}\frac{E^3_\gamma}{\sqrt{s}}
- \frac{2}{3}\frac{m^2_\tau E^2_\gamma}{s}\Bigr)}{[(s-M^2_Z)^2+M^2_Z\Gamma^2_Z](s+2\sqrt{s}E_\gamma-M_H^2)^2},\nonumber\\
F_3(s, E_\gamma, \cos\theta_\gamma)&\equiv&
\frac{\Bigl(-1-\frac{4}{\sin^2\theta_\gamma}+\frac{2\sqrt{s}}{E_\gamma
\sin^2\theta_\gamma}+\frac{2\sqrt{s}E_\gamma}{M^2_Z}-\frac{6m^2_\tau}{\sqrt{s}E_\gamma
\sin^2\theta_\gamma}\Bigr)}{[(s-M^2_Z)^2+M^2_Z\Gamma^2_Z](s+2\sqrt{s}E_\gamma-M_H^2)},\\
F_4(s, E_\gamma, \cos\theta_\gamma)&\equiv&
\frac{\Bigl(-1-\frac{2}{\sin^2\theta_\gamma}-\frac{2\sqrt{s}E_\gamma}{M^2_Z}+\frac{4\sqrt{s}E_\gamma}{M^2_Z
\sin^2\theta_\gamma}+\frac{2sE^2_\gamma}{M^4_Z}\Bigr)}
{[(s-M^2_Z)^2+M^2_Z\Gamma^2_Z](s+2\sqrt{s}E_\gamma-M_H^2)},\nonumber\\
F_5(s, E_\gamma, \cos\theta_\gamma)&\equiv& \frac{
\Bigl[(4-\sin^2\theta_\gamma)\sqrt{s}-2E_\gamma\sin^2\theta_\gamma\Bigr]}
{[(s-M^2_Z)^2+M^2_Z\Gamma^2_Z](\sqrt{s}\sin^2\theta_\gamma)},\nonumber
\end{eqnarray}

\noindent while the coefficients $C_{1,2,3}$ are given by

\begin{eqnarray}
C_1(x_W)&\equiv& \frac{(1-4x_W+8x^2_W)}{x^3_W(1-x_W)^3},\nonumber\\
C_2(x_W)&\equiv& \frac{(1-4x_W)(1-4x_W+8x^2_W)}{x^{5/2}_W(1-x_W)^{5/2}},\\
C_3(x_W)&\equiv& \frac{(1-4x_W+8x^2_W)^2
}{x^2_W(1-x_W)^2},\nonumber
\end{eqnarray}

\noindent where $x_W\equiv \sin^2\theta_W$. We have used the SM
prediction for the $H\tau^+\tau^-$ vertex
$\frac{-iem_\tau}{2\sin\theta_WM_W}$ in order to get the cross
section given in Eq (5). The coefficients $C_{1, 2, 3}$ come from
the expressions for $g_Z$ and the leptonic couplings to the $Z$
and $H$ bosons. The functions $F_{1, 2}$ come from the anomalous
vertex diagrams of Fig. $1(a)$, while $F_5$ comes from the Feynman
diagrams shown in Fig. $1(b, c)$ and $F_{3,4}$ arise from the
respective interference contribution.

\section{RESULTS AND CONCLUSIONS}

In practice, detector geometry imposes a cut on the photon polar
angle with respect to the electron direction, and further cuts
must be applied on the photon energy and minimum opening angle
between the photon and tau in order to suppress the background
from tau decay products. In order to evaluate the integral of the
total cross section as a function of the parameters
$h^{Z\gamma}_1$ and $h^{Z\gamma}_2$, we require cuts on the photon
angle and energy to avoid divergences when the integral is
evaluated at the important intervals of each experiment. We
integrate over $\cos\theta_\gamma$ from $-0.74$ to $0.74$ and
$E_\gamma$ from 5 $GeV$ to 45.5 $GeV$ for various fixed values of
the mass $M_H$. These cuts on the photon energy and polar angle
were used by the L3 Collaboration in order to reduce the
contribution coming from initial-state radiation (ISR).
Accordingly, we did not include in our calculation of the cross
section the contribution due to ISR. Using the numerical values
$\alpha=1/137.03$, $\sin^2\theta_W=0.2314$, $M_{Z_1}=91.18$ $GeV$,
$\Gamma_{Z_1}=2.49$ $GeV$ and $m_\tau= 1.776$ $GeV$, we obtain the
cross section $\sigma=\sigma(h^{Z\gamma}_1, h^{Z\gamma}_2, M_H)$.
As was discussed in Ref. \cite{L3}, $N\approx\sigma(h^{Z\gamma}_1,
h^{Z\gamma}_2, M_H$), and using Poisson statistics
\cite{L3,Barnett}, we require that $N\approx\sigma(h^{Z\gamma}_1,
h^{Z\gamma}_2, M_H$) be less than 1559, with ${\cal L}= 100$
$pb^{-1}$, according to the data reported by the L3 collaboration
\cite{L3}. A similar number of events was obtained for the same
process by the OPAL Collaboration \cite{OPAL}. The experimental
value obtained by the L3 Collaboration for the cross section and
the respective branching ratio of the $H \to Z\gamma$ decay are
given by $\sigma(e^+e^-\to \tau^+ \tau^- \gamma)=(1.472\pm
0.006\pm 0.020)$\hspace{0.5mm}$nb$ \cite{L3,Taylor} and $Br(H \to
Z\gamma)< 10^{-3}$ \cite{Data}. Taking this into consideration, we
get limits on $h^{Z\gamma}_1$ and $h^{Z\gamma}_2$ as a function of
$M_H$. For example, for a Higgs boson mass $M_H=130$ $GeV$ we get
the limits \cite{ICHEP2010}

\begin{eqnarray}
&&|h^{Z\gamma}_1|<0.047,\nonumber\\
&&|h^{Z\gamma}_2|<0.081,
\end{eqnarray}

\noindent while for $M_H=145$ $GeV$ we get

\begin{eqnarray}
&&|h^{Z\gamma}_1|<0.11,\nonumber\\
&&|h^{Z\gamma}_2|<0.19.
\end{eqnarray}

We plot the total cross section in Fig. 2 as a function of the
Higgs boson mass $M_H$ for the values $h^{Z\gamma}_1=0.047$,
$h^{Z\gamma}_2=0.081$ and $h^{Z\gamma}_1=0.042$,
$h^{Z\gamma}_2=0.045$ given in Eqs. (8) and (9). We observe in
this figure that the cross section of the process $e^+e^- \to
\tau^+\tau^-\gamma$ decreases with the increase of the Higgs boson
mass $M_H$. In Fig. 3 we show the region excluded at $95 \%$ C. L.
for the branching ratio BR($H\to Z\gamma$) using Eq. (4) and our
limits obtained for the coupling $h^{Z\gamma}_1$. We notice  an
improvement of about an order of magnitude with respect the
results obtained by the L3 Collaboration from the process
$e^+e^-\to H\gamma$ \cite{Achard}.

In conclusion, we have analyzed the constraints imposed on the
$HZ\gamma$ coupling from the known data for the process $e^+e^-\to
\tau^+ \tau^- \gamma$ obtained by the L3 Collaboration \cite{L3}.
We have made similar analysis using LEP data in order to improve
previous limits on the $ZZ\gamma$ and $Z\gamma\gamma$ vertices
\cite{Perez1,Gutierrez}, the magnetic and electric dipole moments
of tau neutrinos \cite{Maya} and the tau lepton \cite{Gutierrez1},
as well as some of the parameters involved in L-R symmetric and
$E_6$ superstring model \cite{Gutierrez2}. In the present case,
our bounds shown in Eqs. (8, 9) are close to the limits expected
in the annihilation process $e^+e^-\to HZ$ with polarized beams
\cite{Rindani}, and an order of magnitude better than the limits
obtained for the same process by the L3 Collaboration
\cite{Achard}. In particular, we were able to improve the bounds
on the $HZ\gamma$ vertex because we did not need to use in our
analysis the partial decay rates of the Higgs boson used in Ref.
\cite{Achard}.

\vspace{1.5cm}

\begin{center}
{\bf Acknowledgments}
\end{center}

We acknowledge support of RedFAE, CONACyT and SNI (M\'exico).

\vspace{1cm}


\newpage

\begin{figure}[t]
\centerline{\scalebox{0.7}{\includegraphics{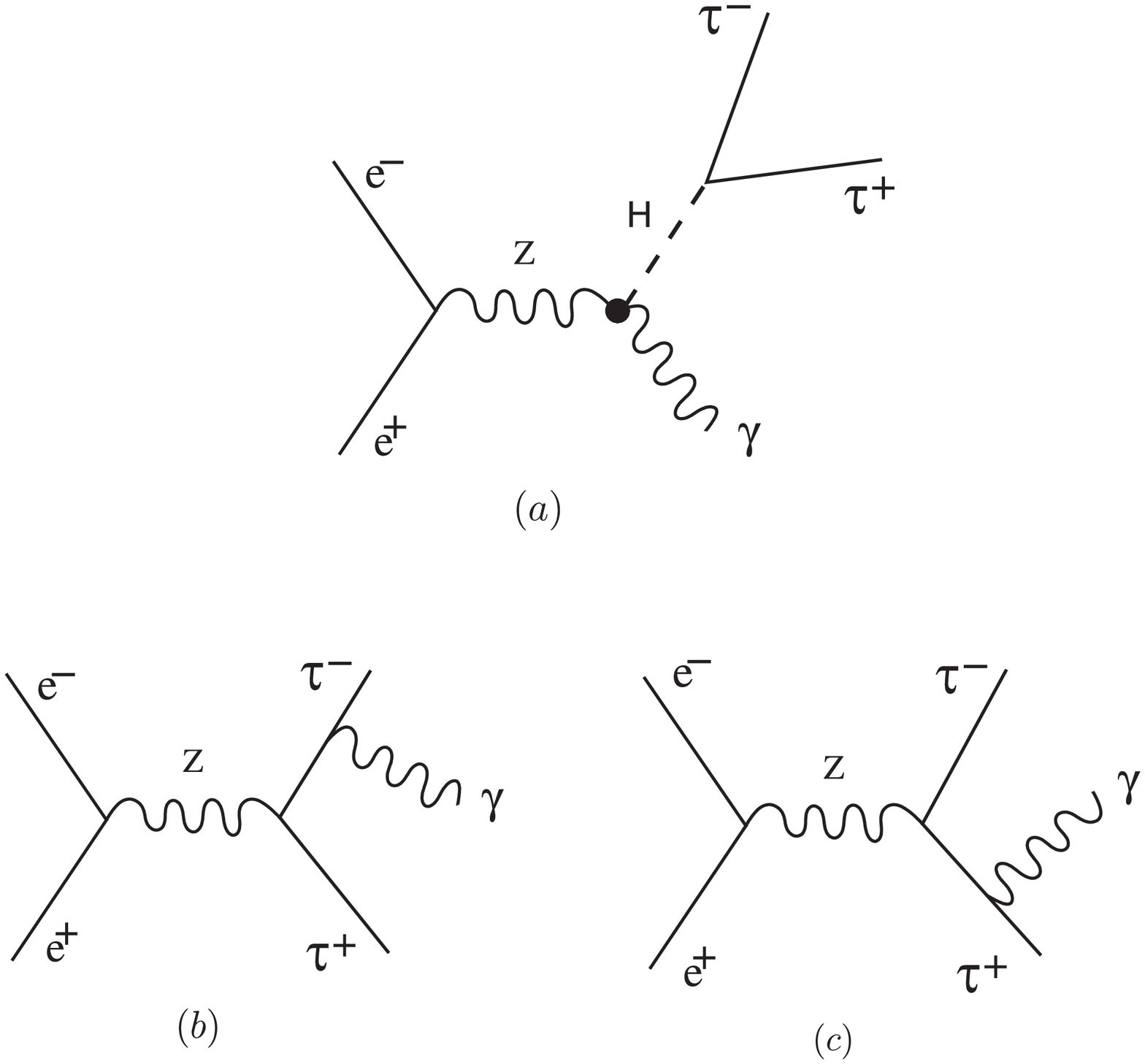}}}
\caption{ \label{fig:gamma} Feynman diagrams for the process
$e^+e^-\to \tau^+ \tau^- \gamma$ induced by the anomalous vertex
$HZ\gamma$ (a) and the SM (b, c) when the $Z$ vector boson is
produced on mass-shell. Diagrams for initial-state radiation are
not considered in the calculation.}
\end{figure}

\begin{figure}[t]
\centerline{\scalebox{1.3}{\includegraphics{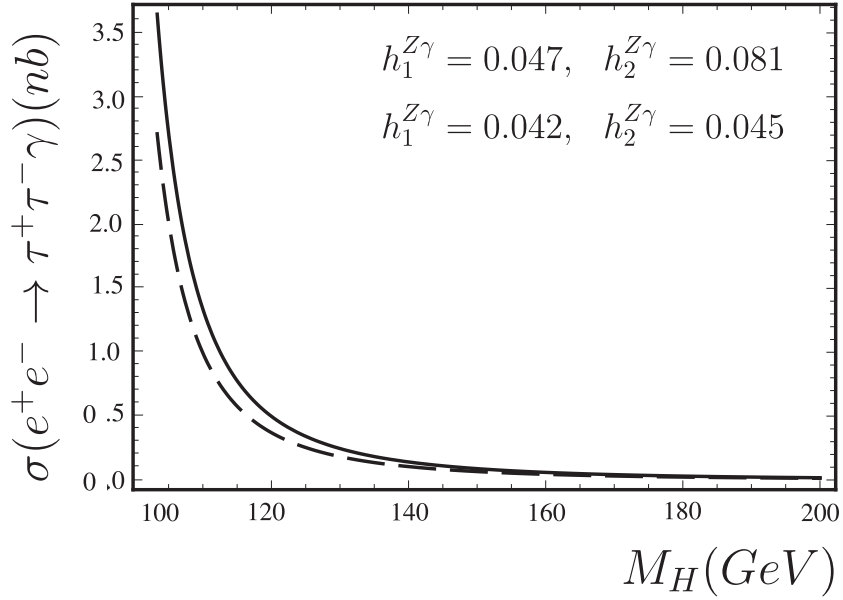}}}
\caption{ \label{fig:gamma} Cross-section for the process
$e^+e^-\to \tau^+\tau^- \gamma$ as a function of $M_H$ with
$h^{Z\gamma}_1=0.047$, $h^{Z\gamma}_2=0.081$ (continuous line) and
$h^{Z\gamma}_1=0.042$, $h^{Z\gamma}_2=0.045$ (dashed line).}
\end{figure}

\begin{figure}[t]
\centerline{\scalebox{1.2}{\includegraphics{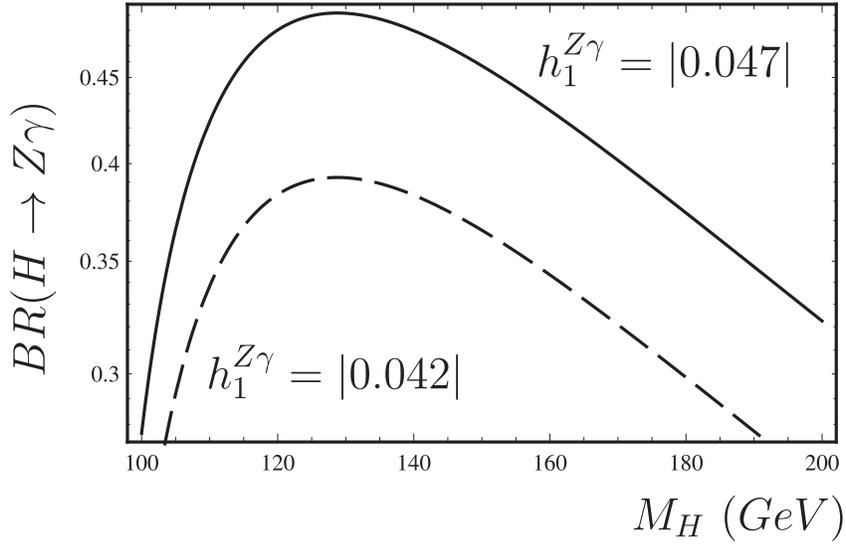}}}
\caption{ \label{fig:gamma} The region above the line is excluded
for the branching ratio BR($H \to Z\gamma$) at $95 \%$ C. L. for a
Higgs mass of $M_H=130$ $GeV$ (continuous line) and $M_H=115$
$GeV$ (dashed line).}
\end{figure}

\end{document}